\documentclass[twocolumn,twoside,slac]{revtex4}
\usepackage{graphicx}
\usepackage{fancyhdr}
\begin{document}

\title{Parallel Reconstruction of CLEO III Data}

%

\author{G.J. Sharp}
\author{C.D. Jones}
\affiliation{Wilson Synchrotron Lab, Cornell University, Ithaca, NY 14853, USA}

\begin{abstract}
Reconstruction of one run of CLEO III raw data can take up to 9 days to complete
using a single processor.
This is an administrative nightmare, and even minor failures result in
reprocessing the entire run, which wastes time, money and CPU power.
We leveraged the ability of the CLEO III software infrastructure
to read and write multiple file formats to perform reconstruction of a single
run using several CPUs in parallel.
Using the Sun Grid Engine and some Perl scripts, we assign roughly equal-sized
chunks of events to different CPUs.
The Raw data are read from an Objectivity/DB database, but the reconstruction
output is written to a temporary file, not the database.
This takes about 6 hours.
Once all the chunks have been analyzed, they are gathered together
in event-number order and injected into Objectivity/DB.
This process takes an additional 6 to 18 hours, depending on run size.
A web-based monitoring tool displays the status of reconstruction.
Many benefits accrue from this process, including
a dramatic increase in efficiency,
a 20\% increase in luminosity processed per week,
more predictable and manageable processor farm load,
reduction in the time to stop the processor farm from up to 9 days
to less than 24 hours,
superior fault tolerance,
quicker feedback and repair times for bugs in the reconstruction code, and
faster turn-around of early runs for data quality and software
correctness checks.
\end{abstract}

\maketitle

\thispagestyle{fancy}

\section{INTRODUCTION}
CLEO III ~\cite{cleo-ref} raw data is reconstructed using the tools provided
by the
CLEO III software infrastructure ~\cite{storage-helpers-ref}~\cite{rapid-ref}.
All the raw data runs for a dataset
are collected and then processed collectively on a compute farm comprising
133 UltraSPARC$^{TM}$ Netra computers.
CPU allocation is performed using Sun Grid Engine (SGE)
~\cite{sge-ref,sge-linux-ref}.
Raw data are read from Objectivity/DB$^{TM}$ ~\cite{objectivity-ref}
and the resulting reconstructed data are written back to the same database.
Since the data are strictly ``write once,
read many'', the storage methods attempt to optimize for read performance.
One of the design decisions that flowed from this was to store the events
in event-number order.

The first implementation of the reconstruction software processed an entire run
sequentially using a single CPU.
A histogram file for monitoring and quality checking was produced.
The reconstructed data was written directly to the Objectivity/DB database.
Many runs were processed in parallel by assigning one run to each of the
available reconstruction farm CPUs.
However, a large run of about 250K events could take up to 9 days to process.

This system had a large number of serious deficiencies:

\def\theenumi{\alph{enumi}}
\begin{enumerate}
\item
It took a long time to stop the processor farm gracefully.
Many CPUs were idle for long periods while everything was halted.
Scheduling periodic maintenance or stopping the farm for emergency
maintenance was fraught with difficulty.
\item
Load-balancing the various CPU farms was difficult.
Once a CPU was loaned to the reconstruction farm it might be committed
for 9 days,
or it might be returned in just 2 or 3 days if it was assigned a short
run, or there was a software failure.
\item
Database locks were held for up to 9 days.
With more than 100 runs being processed at once,
there was a heavy load on the database lock server, 
which hurt performance and interfered with database
administration activities.
\item
Reconstruction averaged about 1.5 to 2 seconds per event.
This resulted in a very low write rate to the database.
Caching efficiencies were lost.
\item
Any failures left the database in an invalid state.
Scarce database administrator (DBA) time was required to clean out the partial
results.
If failures went undetected for some time, an attempt may have been made
to use the corrupted database, resulting in more wasted time, computing
and human resources.
\item
There was a very large window of opportunity for failures to occur while 
the database was open.
For example, power failure, CPU failure or reconstruction 
software bugs resulted in lost CPU time, extra operator intervention 
and DBA intervention.
\item
When starting a new dataset, it is necessary to check the output of
a few (short) runs to ensure that the software and the constants are combining
to produce reasonable output.
If a problem occurs, the fault must be corrected and processing of the entire
dataset restarted.
This took from several days to a week.
Waiting several days for a few short runs to complete so that the output
could be checked led to a substantial waste of resources.
\item
Debugging the reconstruction software components was painful.
It could take days to rerun a job to isolate the bug, fix it and test the fix.
Then the entire run had to be reprocessed to inject the data into the database.
\item
Failed jobs were not automatically restarted because the database
clean-up required human intervention.
For transient failures, automatic
restart would improve resource utilization, reduce operator intervention
and potentially reduce the time taken to complete a dataset.
\end{enumerate}

This litany is by no means the complete list of problems.
These were the problems that needed to be addressed urgently.

\section{THE SOLUTION}
Reducing the wall-clock time to analyze an entire run would clearly resolve
or ameliorate problems a, b, c, g and h.
It was decided to leverage the experience gained from Nile
~\cite{nile-ref, nilebib-ref} to use more parallelism.
Each run could be split into several chunks,
and each chunk assigned to a separate CPU.
However, splitting a run into modest-sized chunks of events and processing
the chunks in parallel potentially conflicts with writing the events out in
event-number order.
Writing the data directly to the database in event-number order would be almost
impossible, and would not address problems e and f.
Fortunately, the CLEO III software infrastructure
was designed to handle different file formats simultaneously.
Therefore each run is split into chunks but the results are not written
directly to the database.
Instead, an efficient binary format is used to temporarily
store the reconstructed data.
Within each chunk, the intermediate output is still in event-number order.
Once all the chunks are reconstructed,
a single job injects the reconstructed data into the database in event-number
order by collating the intermediate files in the correct order.
This process is limited by the write performance of the database.
Now problems c, d, e and f have been dealt with.
Extra fault tolerance is gained because now the failures in reconstruction are
no longer failure modes for updating the database.
No database updates occur until the reconstruction is complete.
An extra benefit is that we are assured that the reconstruction data can
be read and is moderately well-formed.

To further improve fault tolerance, a ``job manager'' process is created for each run.
It splits the run up into chunks and submits a reconstruction subjob for each
chunk.
All the subjobs run in parallel.
Each subjob produces an intermediate output file and a histogram file.
The job manager process monitors the log files of the reconstruction subjobs
to determine when they are all complete.
If any fail it will attempt to restart them, unless it determines that the
failure is permanent (rather than due to a transient problem).
Once all reconstruction subjobs complete it starts the collation
of the intermediate files into the database and merges the histograms.
In principle, these two jobs can occur in parallel.

Web pages showing the status and statistics of completed runs and runs
in process are produced automatically.
In addition to vital statistics about each run
(such as event count, subjob state, beam energy, luminosity and cross-sections),
they contain reasonably accurate
predictions of the completion time for each process.
The predictions are based on the number of events remaining and the
average time per event so far for a subjob or a run.
They also contain a prediction for the completion time of the entire dataset.
(The prediction extrapolates the average processing time per event for
the runs so far processed to the runs awaiting processing.)
This is used to ensure that the data for the next dataset to be processed
are staged in a timely fashion so that the time delay between the completion
of one dataset and the start of the next is minimal.
In fact, we have even started processing a new dataset while the last
few runs of the previous dataset complete.
Any web page entries that have suspicious values are colored in orange or
red to attract the attention of the operator and senior physicists.

\section{THE IMPLEMENTATION}
There are many utilities to simplify the work of the operator.
Only the core of the implementation is described here.

The system is mostly written in object-oriented Perl ~\cite{perl-ref}.
A small number of shell scripts are used to start the reconstruction and
collation software.
The software is not general purpose.
It is closely tailored to the CLEO III environment.
However, the ideas behind it should be applicable to many other domains.

Each run is managed by a job manager process.
When the raw data for a run has been staged to disk,
the operator starts the job manager process for that run.
A submission script is used to start the job managers for pre-staged runs. 
When the SGE job queue is short enough the submission script submits more runs,
one at a time, until the queue is considered long enough.

The job manager does not do any processing of Raw data. 
Instead it accepts four arguments:

\def\theenumi{\arabic{enumi}}
\begin{enumerate}
\item run number
\item reconstruction script
\item collation script
\item histogram merge script
\end{enumerate}

The three scripts are responsible for all data analysis activities. 
The job manager performs the following tasks:

\begin{itemize}
\item
It splits its run into roughly equal-sized chunks.
We chose approximately 20K events so that reconstruction subjobs would finish
in about six hours.
\item
It uses configuration files to estimate the approximate output volume for each
script and selects a disk for the output of each reconstruction 
subjob and collation subjob.
The output disk selection system attempts to avoid overfilling any output
disks and overloading any of the output disk servers.
If necessary, a job is delayed until a server with sufficient capacity
to handle it becomes available.
The histogram output is quite small and
is stored in a fixed directory for each dataset.
\item
It starts a subjob (using SGE) to reconstruct that run,
writing the output to an intermediate file.
\item
It waits for all the reconstruction subjobs to complete.
\item
If any fail due to transient problems it attempts to 
rerun the failed jobs.
Any permanent failures result in email to the operator.
\item
Once all the reconstruction subjobs complete successfully the 
job manager starts the collation subjob.
This takes from 6 to 18 hours, depending on the size of the run.
If it fails, the job manager sends email to the operator and terminates.
It cannot be rerun automatically because 
the DBA must clean out the mess left by the failed attempt.
\item
Once the collation completes successfully, the histograms are
merged using PAW ~\cite{paw-ref}.
This takes between 5 and 15 seconds.
It could be done in parallel with the collation subjob,
but the extra code required for the parallelism is not worth the complexity.
\end{itemize}

Note that the job manager can be restarted if it is killed or if the CPU 
where it is running fails.
When started it detects any subjobs that are still running,
detects any subjobs that have completed while it was gone,
and then resumes where it left off.
If the job manager detects any failures it sends email to the operator,
so the operator does not need to perform as much system monitoring.
The automatic retrying of jobs also reduces operator intervention and ensures 
timely completion of jobs.
It can result in wasted CPU power if it is retrying jobs with permanent
failures, but the waste is slight compared with the benefits.

\begin{figure*}[t]
\centering
\includegraphics[width=135mm]{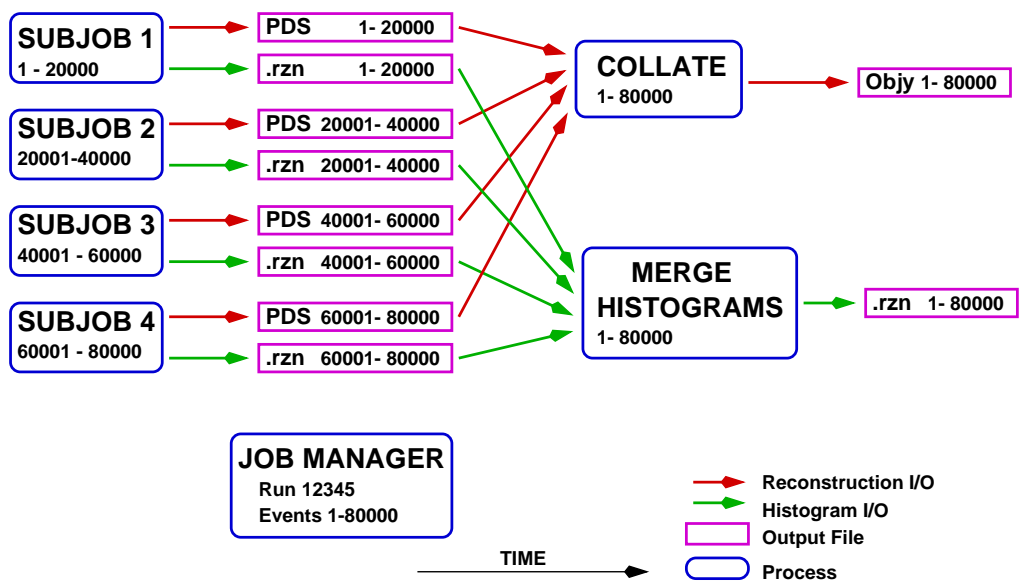}
\caption{Block Diagram of Processes and I/O.
Time increases from left to right.
The rectangular boxes indicate outputs.
The boxes with rounded corners indicate processes.
The job manager can be (re)started anywhere in the life of the reconstruction
of a run.
Merge and Collation may be run in parallel, but the short life of the merge
subjob renders this a minor optimization.
} \label{procview-f1}
\end{figure*}

Figure~\ref{procview-f1}) gives an overview of the process as time increases
from left to right.
The output file marked ``PDS'' is the intermediate file. The output files
labeled ``.rzn'' are the histogram files.

The reconstruction command passed to the job manager can be any executable.
For CLEO III it is a shell script that runs the appropriate CLEO III software
for reconstruction.
Likewise for the collation subjob.
The merge histograms subjob is also a shell script, but it runs a simple
PAW program that merges the histograms produced by the reconstruction subjobs.

\section{RESULTS}

We selected a chunk size such that it takes 
about 6 hours for the reconstruction phase.
The amount of time for a job can be predicted more accurately after about 10\%
of the events are done.
An estimate can also be made for the collation job.
The estimates are calculated and displayed on the status web pages.
This greatly assists in load balancing and the shorter execution time greatly
increases the flexibility for transferring 
CPUs from one farm to another to meet spikes in demand.

The total run time for a job has been reduced to 12-24 hours. The long 
time required to inject the data into the Objectivity/DB database is due, in 
part, to the serial injection of each event. No mechanism has been devised 
for injecting the data in parallel. Implementation issues in Objectivity/DB 
make it undesirable to have a separate database for each chunk - database 
identifiers would run out too quickly.

Fault tolerance has been dramatically improved.

\def\theenumi{\alph{enumi}}
\begin{enumerate}
\item
If one event triggers a failure in the reconstruction software, only
the events in that small chunk need to be reprocessed instead of the entire
run.
\item
Bugs in reconstruction do not result in a run partially populated into
the Objectivity/DB, which reduces the administrative overhead of cleaning
out partial runs.
Failures in collation still occur, but far less frequently.
\item
Because there is a process monitoring the progress of each run much
of the fault detection and some recovery is automated.
\end{enumerate}

Reviewing the initial problems described in the introduction:

\begin{enumerate}
\item
The time to stop the processor farm has been reduced to between 6 and 
18 hours, depending on the size of the collation subjobs that are running.
Much of the time, few collation jobs are running, and a decision could be made
to stop the farm almost immediately and clean out the handful of runs in
collation phase.
Reconstruction jobs can be stopped at any time, since restarting the job manager
at a later time will cause the interrupted jobs to be automatically rerun.
\item
The relatively short runtime for each subjob,
and the completion time estimates for each subjob greatly simplify the CPU
farm load balancing issues.
\item
Database locks are now held only during collation, which is the minimum 
possible time for holding the locks.
\item
The write rate to the database is now about 3 to 4 events per second, 
compared with several seconds per event under the old system.
The performance is disappointing, but as good as we can do with the current
database schema.
\item
The only failure modes during writing the database are hardware 
problems and collation failures.
The reconstruction failures have been eliminated.
\item
The window for failures (especially hardware and power failures) has 
been reduced from 9 days to 18-24 hours.
This is a dramatic improvement.
\item
The potential CPU waste at the start of a new dataset when checking
the reasonableness of the reconstruction output for the first few runs of
a dataset has been reduced to substantially lower than 24 hours, and as little
as 12 hours if we use a small run.
\item
Reproducing bugs is much easier and quicker.
\item
The automatic rerunning of failed reconstruction jobs often results 
in quicker processing of a dataset.
It also aids in debugging, since a second attempt is quickly made,
which will indicate if a fault is transient, pseudo-random or reproducible.
\end{enumerate}

Some potential disadvantages of the new system are worth noting.
\begin{itemize}
\item
The data delivery rate to the Objectivity/DB disk servers is much higher
but for shorter periods.
In pathogenic situations this may result in performance degradation.
Normally it is not a problem, since only a few collation
jobs are running at the same time, and each may be assigned a different disk
server for its output.
However, since the collation subjob tends to run for
much longer than the reconstruction subjob, towards the end of a dataset
it is possible that a large number of collation jobs are running simultaneously.
Because the output disk selection system prevents overloading a server, it
may be the case that some collation jobs are delayed until a server has
sufficient capacity to handle the job.
\item
The total number of CPU cycles used by this method of running reconstruction
is greater than the old system. The compensation is that the wall-clock time
to complete a dataset is much less.
By consuming slightly more CPU cycles
to do the work, far fewer CPU cycles are wasted or idle.
\item
There are more log files, requiring somewhat more disk space.
This is not a significant burden.
\item
The operator frequently gets bored and drifts off to
refill printers or load tapes.
\end{itemize}

The rate of reconstruction has increased about 20\% due to the introduction
of finer-grained parallelism.
Several factors led to this.
The most significant is having accurate estimates for the completion time
of a dataset.
This allows data for the next dataset to be processed to be staged to
disk from tape ``just in time''.
Several bugs have been fixed as a result of the easier debugging,
so there are fewer crashes.
Automatic submission of new jobs when the job queue is almost empty
also ensures almost full utilization of the available CPU power.

\section{FUTURE DEVELOPMENTS}

The next step in the automation of reconstruction is to automatically stage 
raw data from the HSM file systems to cache disk and then feed the staging 
information directly to the reconstruction system.
In this way, newly staged runs are almost immediately available for processing.
No operator intervention will be required for them to be reconstructed.
Automatically staging the reconstruction 
output back to the HSM file system is the final step in the process.

The SGE is available on Linux ~\cite{sge-linux-ref}.
Both the CLEO III software infrastructure and Objectivity/DB are being ported
to Linux.
Once these ports are available we hope to be able to run CLEO III
reconstruction on low-cost Linux nodes, which should yield substantially
better price/performance.

Major performance gains will be made by replacing Objectivity/DB with a
simpler and faster database management system that allows dramatically
faster storage of the reconstruction output.

\section{CONCLUSIONS}

The reconstruction system described above would not have been possible
without support for multiple input and output formats in the CLEO III software
infrastructure.
The flexibility it provides has been of great benefit here
and in other respects.

It is true that more CPU cycles are used to process the reconstruction 
than in the old system.
However, since far fewer cycles are wasted, and CPUs 
are idle for a much smaller percentage of the time, the CPU utilization is 
much higher, and the time to complete a dataset has decreased.
The extra cost of the fault tolerance is a worthwhile trade-off.

Accurately predicting the completion time for processing of a dataset
substantially reduces delays between the completion of one dataset and
the start of the next.

Quantizing the time for each job greatly simplifies system management.
Predictable behavior brings huge benefits.

Automating many of the tasks catches more problems and catches them much
earlier than human searching of log files.
Since there are hundreds of log files to check on,
and the computer is well suited to the task, reducing
human involvement is worthwhile.
The operator has fewer details to attend to,
is more productive and less stressed as a result.
He is most grateful.

The fault-tolerance features demonstrated their utility when Ithaca experienced 
an ice storm in the first week of January 2003. After cleaning out the database 
for the 4 or 5 collation jobs killed by the power failure,
the job managers were restarted.
All the failed subjobs were correctly rerun automatically.
Almost no time was spent checking log files and cleaning up the entrails.
That freed staff to focus on other recovery matters and lightened the workload
in a time of crisis.
Under the old system, 133 jobs would have had to have been cleaned out of the
database.
The restart could have taken several days to prepare for,
rather than just a few minutes.

\begin{acknowledgments}
The authors would like to thank the following people:
Jean Duboscq and David Kreinick for removing obstacles (as good managers 
should), Dan Riley for excellent technical advice and Curtis Jastremsky
for keeping a positive attitude during testing, and giving us the insights
that only an operator has.
\end{acknowledgments}

\end{document}